# The *"local atomic packing"* of a single-component glass is quasi-crystalline

Farid F. Abraham
IBM Almaden Research Center
San Jose, CA

*Contact Address
865 Paullus Drive
Hollister CA 95023
fadlow@outlook.com

Our earlier Monte Carlo simulations of metastable supercooled-liquid and glass phases of Lennard-Jones atoms found several distinct signatures for identifying the glass transition boundary; i.e., the density, enthalpy, and pair distribution function dependences on temperature and pressure are different for the two phases (F. F. Abraham, J. Chem. Phys., 72, 359 (1980)). In this extension of that study, we base our analysis on the Ackland-Jones (A-J) method for determining the local crystal packing about each atom (G. Ackland & A. Jones, PRB 73, 054104 (2006)). It focuses on the angular distribution of the *local neighborhood of atoms* surrounding each individual atom and compares it with the known FCC, HCP, BCC, and icosahedron packing within a specified "uncertainly" from perfect packing. Remarkably, the A-J method applied to our simulated glass states indicates that the *local atomic packing* about the individual atoms are predominantly "*quasi-crystalline*"; i.e., "quasi" FCC, HCP or BCC.

## INTRODUCTION

By employing the isothermal-isobaric Monte Carlo method of classical statistical mechanics [**1, 2**]**,** we are investigating the structural and thermodynamic features of the supercooled liquid-glass transition region by abruptly cooling and/or compressing an equilibrated simple liquid. In our earlier (37 years ago) simulation studies **[3, 4]**, we reported on the density, enthalpy, and pair distribution function of the metastable states prepared by instantaneously quenching or crushing a Lennard-Jones liquid beyond the liquid-solid phase boundary. We found "kinks" in their linear behavior which we interpreted as defining the supercooled-liquid/glass phase boundary in the *(P, T)* plane. We calculated, from liquid-state perturbation theory, that the equivalent hard-sphere packing fraction of the supercooled liquid at the glass transition is 0.53, irrespective of whether the glass is reached by quenching or crushing. Furthermore, the character of the pair distribution function as a function of the degree of metastability suggested the possibility of a structural short-range order that is indicative of an incomplete FCC packing, the exclusion of certain peak positions being governed by the dispersion of the first shell atoms. In this study, our analysis is based on the Ackland-Jones (A-J) method for determining the local crystal packing about an atom [**5**]. The A-J method focuses on the angular distribution of the atoms surrounding a given atom and compares it with the known body centered cubic "BCC", face centered cubic "FCC", hexagonal close packed "HCP", and icosahedron packing with "a specified allowed uncertainly", i.e., they made a large number of samples in each crystal structure by randomly displaced each particle from its perfect lattice site. They found that the cosines of the angles between the bonds give a clear distinction using angular distribution functions (ADF); these are frequency distributions of angle cosines among the immediate neighbors of a given particle. Remarkably, applying this method to our simulated glass states suggests that the individual *local packings* of the glass atoms are predominately quasi-crystalline.

A discussion of nomenclature is required. Our reference to the A-J's FCC, HCP, BCC and icosahedron assignments as nanocrystallites is certainly counter to convention. Convention states that a nanocrystalline (NC) material is a polycrystalline material with a crystallite size of only a few nanometers [**8**]. These materials fill the gap between amorphous materials without any long range order and conventional coarse-grained materials. Nanocrystalline materials are single- or multi-phase polycrystalline solids with a grain size of a few nanometers. In this study, the nanocrystal can simply be a first two shells of atoms surrounding a core atom while A-J's "other" does not satisfy their criteria for being nanocrystalline.

## THE SIMULATIONS

The (N, P, T) Monte Carlo procedure used in this study has been described by McDonald [**2**]. The interatomic force law was arbitrarily chosen to be Lennard-Jones 12:6 with the well-depth and size parameter set to unity (reduced units). In order to simulate the bulk, the standard periodic boundary conditions were imposed with respect to translations parallel to the faces of the computational cube composed of 1372 atoms. In each simulation or "experiment", we started with a fluid configuration of a well-equilibrated fluid at reduced temperature $T^*=1.0$ and reduced pressure $P^*=1.0$. Then, we abruptly quenched the system to a new temperature $T^* <1.0$ by simply setting the temperature in the Monte Carlo procedure to the new desired value. After "equilibrating" the system with a 50% acceptance ratio, further individual moves were performed to obtain the average density, the enthalpy, and the pair distribution function $g(r)$. Experiments were performed for constant pressure P*=1.0 and various temperature quenches 0. 1< T* <1.0. Extreme care was exercised to guarantee that metastable equilibration was established and maintained over the averaging interval by monitoring the density, enthalpy, and structure statistics. In one case we observed an instability initiated by a nucleation event and crystallization to an imperfect FCC solid.

THE SIMULATION RESULTS AND INTERPRETATION

As illustrated in our Figs. 1a and Fig. 3 in reference [4], the pair distribution functions g(r) show the general features expected. In the liquid and supercooled-liquid state, two smooth peaks exist; a first prominent peak and a second smaller peak corresponding to the first and second coordination shells of an atom in the liquid, respectively. As the instantaneous temperature quenches probes deeper into the metastable region, the first peak becomes more pronounced in magnitude and narrower in width, the first minimum decreases in magnitude, and the second peak gradually flattens in shape with an eventual "bimodal splitting" at very low temperatures (P*=1.0,T*~0.4-0.5). The development of the split second peak indicates a glass atomic packing, this being the principal structural feature in the experimental PDF of amorphous materials that previous theoretical models have attempted to describe. However, the second peak flattening preceding the fully developed splitting may be the "signature" that the glass transition region has been reached.

In our earlier study, we defined an empirical parameter $R^* = g_{min}/ g_{max}$ and it is graphically shown in Fig. 1. Our present simulations for constant pressure P*=1.0 and series of temperature quenches T*= 0.9 to 0.1 in 0.1 increments shows the classic R* behavior found earlier, the kink occurring around T*==0.4.

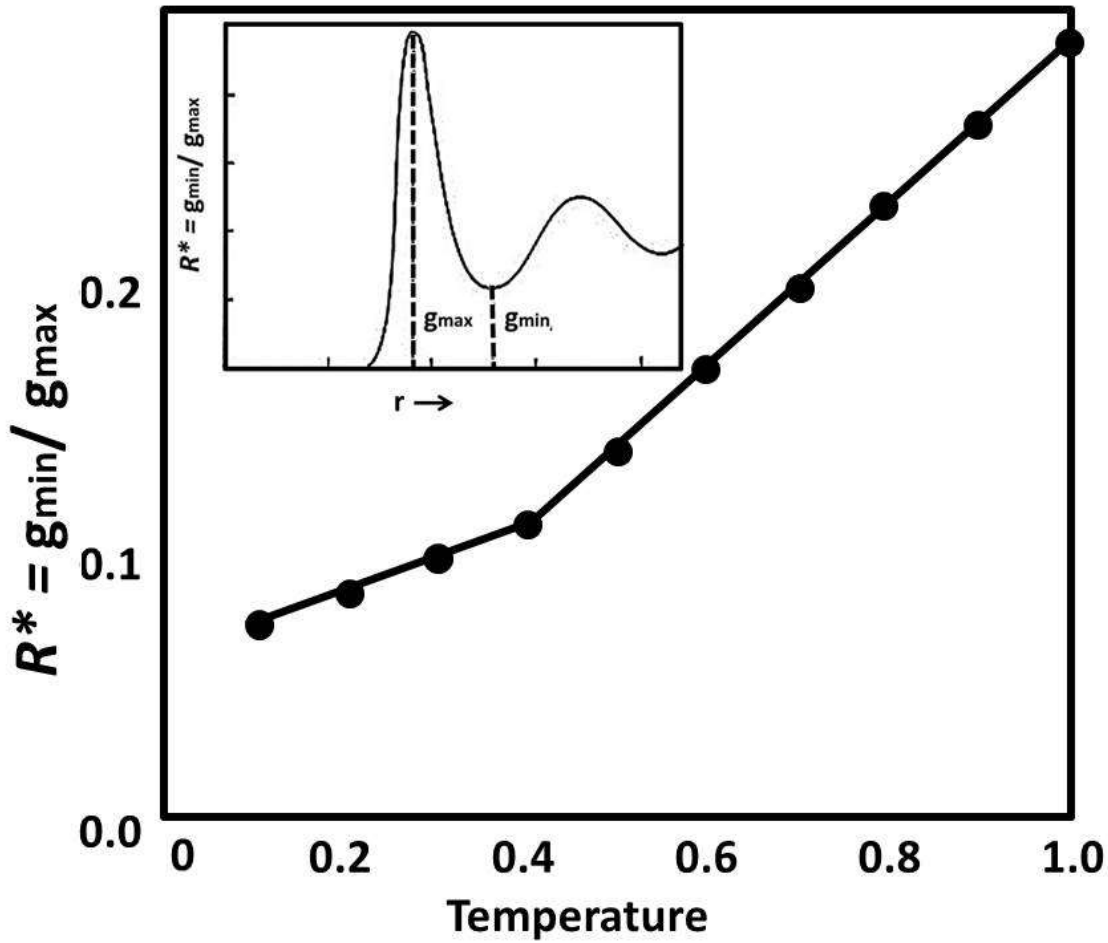

Fig. 1. Insert: Graphical presentation of the definition of the radial distribution function R* parameter. $R^* = g_{min}/g_{max}$ versus temperature for the present simulations. The glass transition temperature "kink" is at T~0.4

We present our analysis is based on the Ackland-Jones (A-J) method for determining the local crystal packing about each atom [**5**]. The method's important aim was to identify imperfections in crystalline packing, certainly not intended for studying the structure of the glass state! It focuses on the angular distribution of the atoms surrounding a given atom and compares it with the known FCC, HCP, BCC, and icosahedron packings (with noise or a specified measure of uncertainty.). If there is no "good" fit, it chooses "other." You might suspect that most of the glass atoms would be "other." We used the Ackland-Jones method implemented in Alexander Stokowski's "OVITO Open Visualization Tool" [6]. Ovito is a graphics package with many modern tools for analyzing atomistic simulations.

In the next figure (Figs. 2), graphics output from the bond-angle analysis is presented for the different temperature quenches. Only the close-packed structures,

FCC (yellow) and HCP (blue) atoms are plotted, and atoms separated by a reduced distance less than 1.25 are connected by a (green) bond. The relative populations of the various A-J types of atoms are presented in Table 1. While the number of atoms and bonds increase with decreasing temperature, interconnectivity of the bonds does not span the computational cell until a temperature T* approximately equal to T*=0.4. This corresponds to the glass transition point determined by the inflection of the R* dependence (Fig. 1). One might question how a glass can be an aggregate of "first-shell crystallites." Because of the added noise in the identification analysis, the local environment reflects such a crystal packing but should not be taken literally. (NOTE: Even where the relative population of HCP is greater than FCC in the disordered packing, this does not reflect the fact that when the disordered system nucleates to the crystal, it will be FCC [7].)

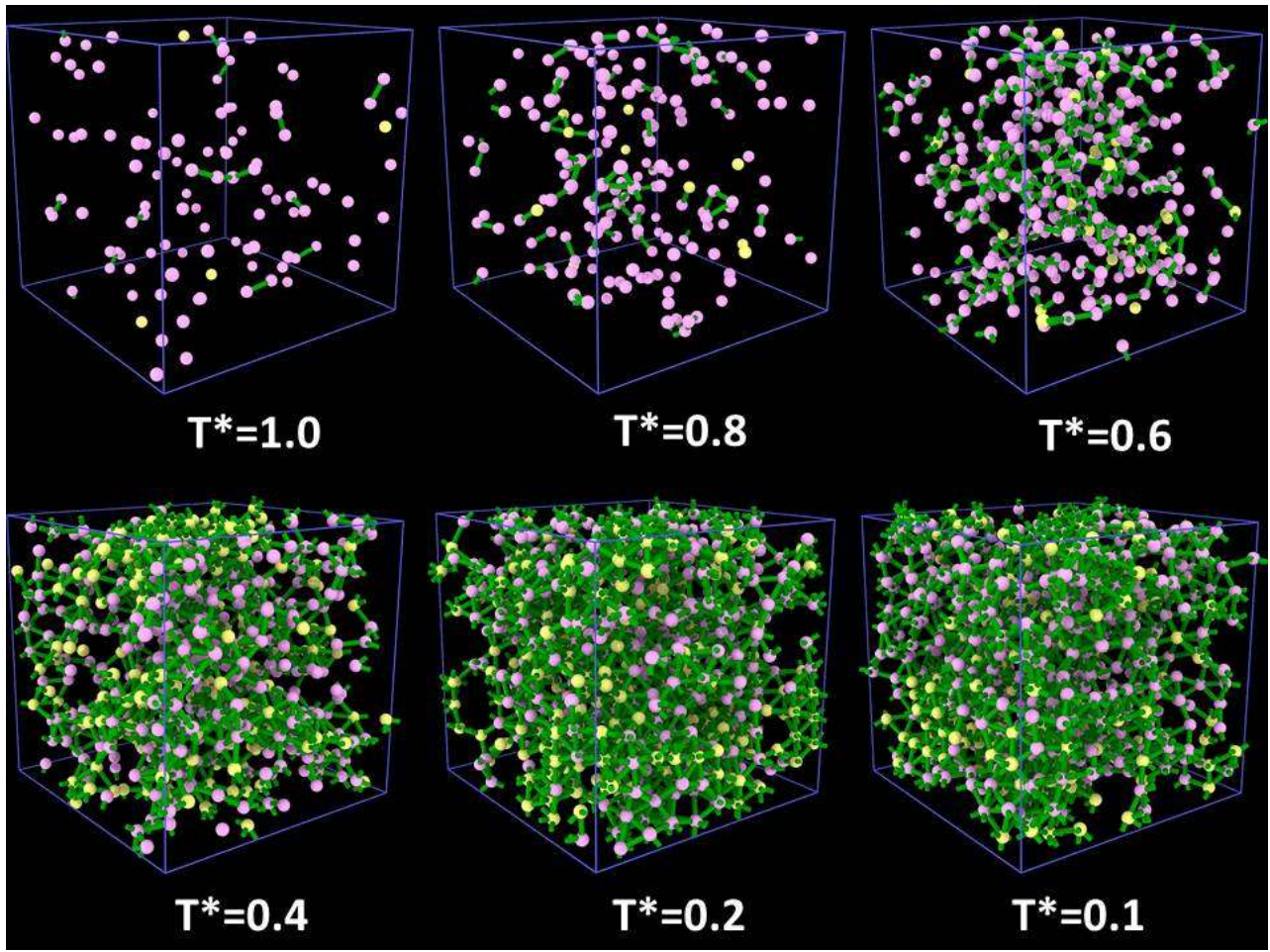

Fig. 2. For the liquid phase, the percentage of A-J crystallites is very small (see Table 1). For the super-cooled liquid (T*=~0.6), the percentage of A-J crystallites grows, but interconnection is low. Cooling to the onset of the glass phase (T*=0.4) gives rise to interconnected bond-chains spanning throughout the computational

cell. The percentage of "Other" is at a lower bound of approximately 20%. Only the close-packed structures, FCC (yellow) and HCP (blue) atoms are plotted, and atoms separated by a reduced distance less than 1.25 are connected by a (green) bond.

| T*=1.0 | | | T*=0.8 | | | T*=0.6 | | |
|---|---|---|---|---|---|---|---|---|
| Name | Count | Fraction | Name | Count | Fraction | Name | Count | Fraction |
| Other | 1187 | 86.5% | Other | 1051 | 76.6% | Other | 731 | 53.3% |
| FCC | 4 | 0.3% | FCC | 10 | 0.7% | FCC | 28 | 2.0% |
| HCP | 102 | 7.4% | HCP | 169 | 12.3% | HCP | 330 | 24.1% |
| BCC | 74 | 5.4% | BCC | 136 | 9.9% | BCC | 260 | 19.0% |
| ICO | 5 | 0.4% | ICO | 6 | 0.4% | ICO | 23 | 1.7% |
| T*=0.4 | | | T*=0.2 | | | T*=0.1 | | |
| Name | Count | Fraction | Name | Count | Fraction | Name | Count | Fraction |
| Other | 424 | 30.9% | Other | 288 | 21.0% | Other | 296 | 21.6% |
| FCC | 98 | 7.1% | FCC | 327 | 23.8% | FCC | 214 | 15.6% |
| HCP | 463 | 33.7% | HCP | 453 | 33.0% | HCP | 534 | 38.9% |
| BCC | 350 | 25.5% | BCC | 293 | 21.4% | BCC | 296 | 21.6% |
| ICO | 37 | 2.7% | ICO | 11 | 0.8% | ICO | 32 | 2.3% |

Table 1. Statistics of the different types of A-J crystallites as a function of temperature.

In Fig. 3, we present a "mirror" representation of Fig. 3. The solid space represents volume occupied by atoms classified as the non-crystalline "other" and the opened regions is where atoms are classified as "crystalline." We note that T*<0.5, significant "porosity" of the cube is seen, denoting the highly interconnect ordered region below the glass transition.

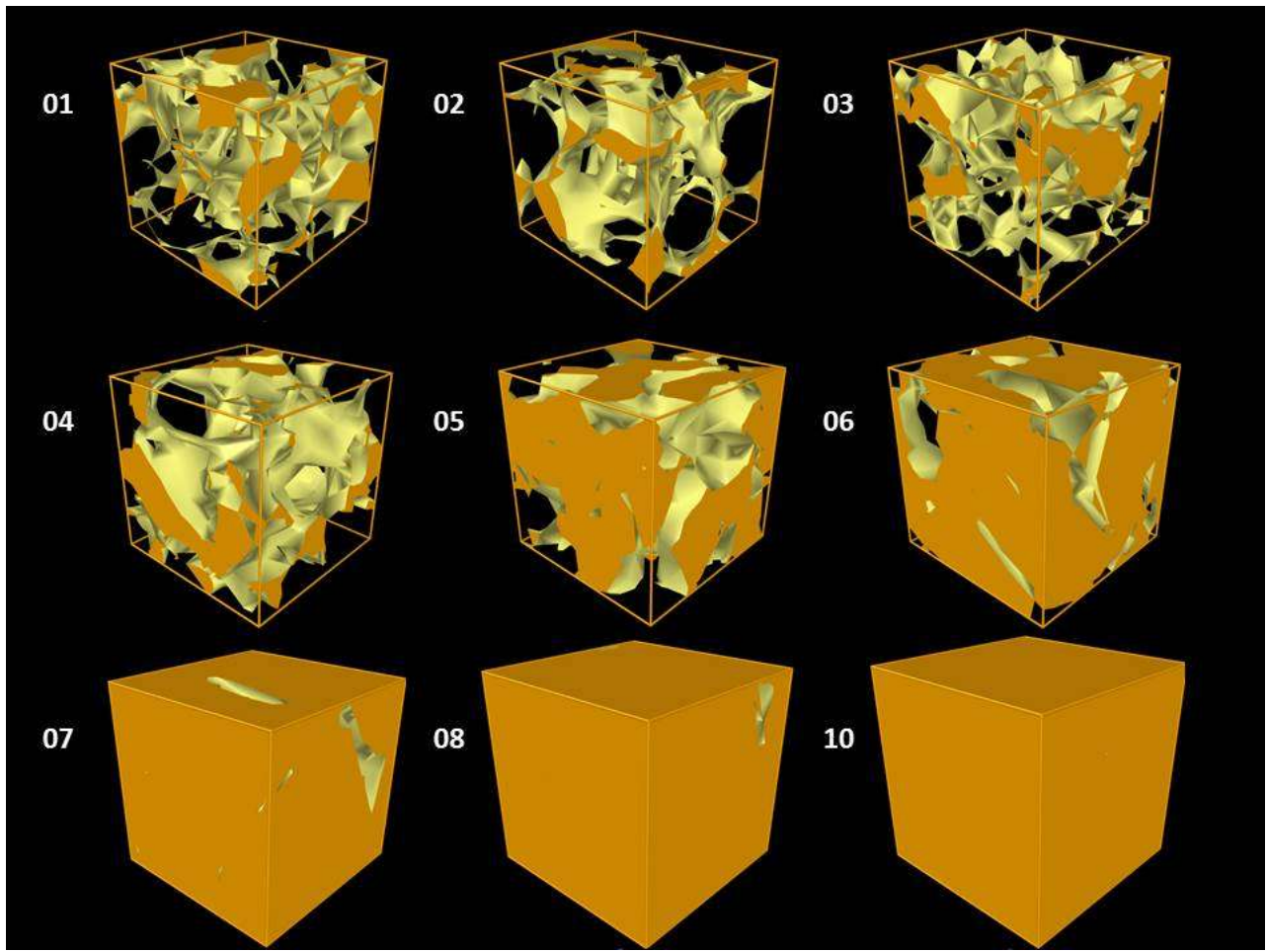

Fig. 3. Mirror representation of Fig. 3 for temperatures 0.1 to 1.0. The solid space represents volume occupied by atoms classified by A-J as non-crystalline "other" and opened regions where atoms are classified as "crystalline."

We conclude that the glass transition region is when long-range interconnectivity between quasi-crystalline A-J atoms is achieved. We can graphically observe this in the following representation. In Fig. 4, we plot the maximum cluster size of interconnected quasi-crystalline atoms and the total number of independent clusters as a function of temperature. We note that as the temperature cools through the glass transition region, the maximum cluster size grows rapidly through the coalescence of smaller clusters leading to a precipitous drop of number of isolated small clusters. We speculate that this is the origin of the onset of the significant increase of viscosity at the glass transition.

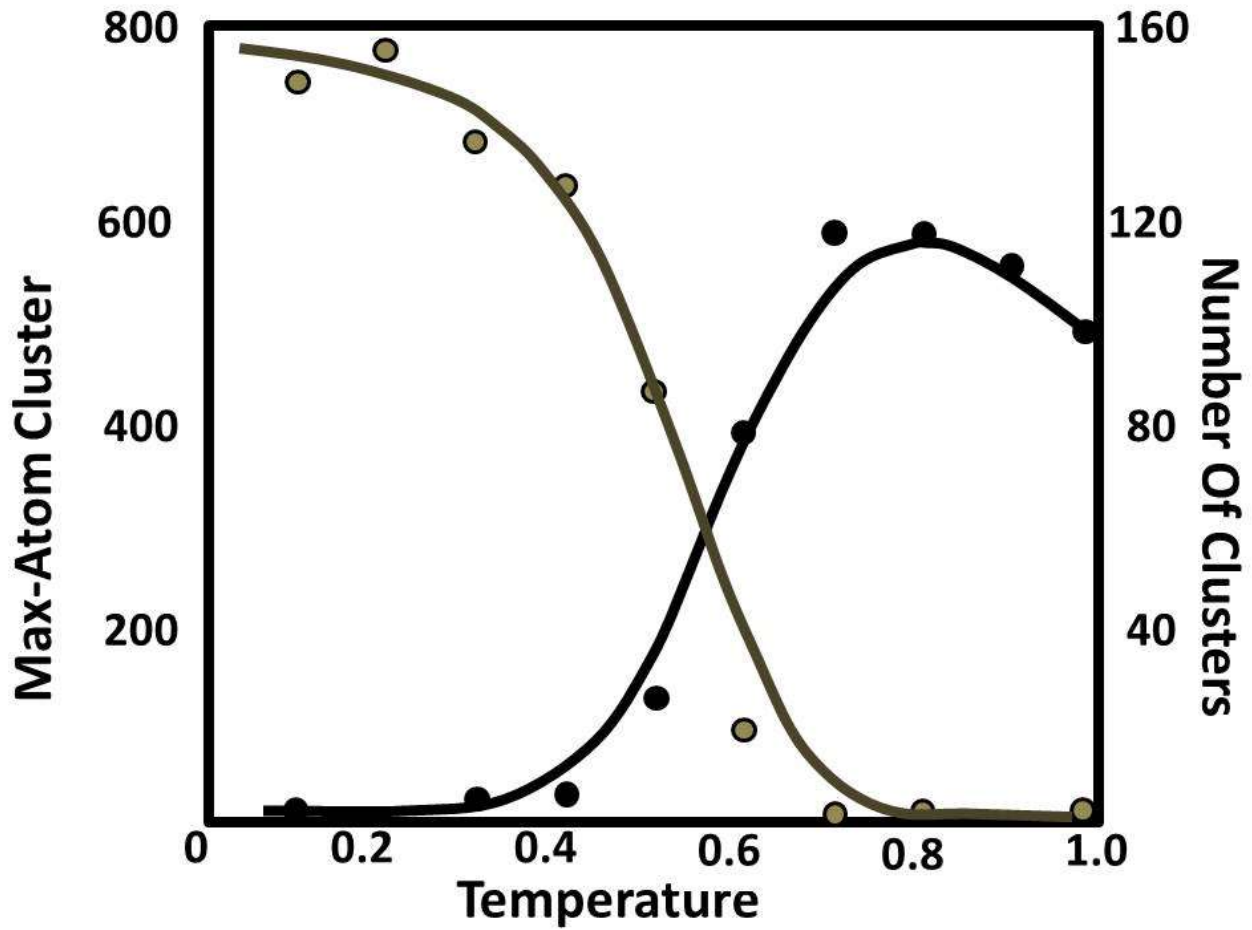

Fig. 4. Maximum interconnected quasi-crystalline A-J atoms and number of clusters as a function of temperature.

**CONCLUSION**

Based on the Ackland-Jones analysis applied to disorder systems as a function of temperature; i.e. the transition from the liquid to glass regions, we have shown that the local packing environment around the individual atoms becomes more crystalline-like and that this local-crystalline structure becomes highly interconnected in the glass region. We believe that this is a significant feature of the glass state.

**Acknowledgement**

I am indebted to Professor A. Stukowski for much help with Ovito.

# THE REFERENCES